# Inconsistencies between the forces from code live load models and real traffic on truss bridges

Alejandro Hernández–Martínez. Civil and Environmental Engineering Department, Engineering Division, Universidad de Guanajuato. Av. Juárez 77, Zona Centro. Guanajuato, Gto., México.
alejandro.hernandez@ugto.mx

Adrián D. García–Soto. Civil and Environmental Engineering Department, Engineering Division, Universidad de Guanajuato. Av. Juárez 77, Zona Centro. Guanajuato, Gto., México. adgarcia@ugto.mx

Hugo Hernández–Barrios. Civil Engineering Faculty. Universidad Michoacana de San Nicolás de Hidalgo. Ciudad Universitaria, Morelia, Mich., Mexico. hugohernandezbarrios@yahoo.com.mx

Jesús G. Valdés–Vázquez. Civil and Environmental Engineering Department, Engineering Division, Universidad de Guanajuato. Av. Juárez 77, Zona Centro. Guanajuato, Gto., México. valdes@ugto.mx

**Abstract**

The forces acting on bridge structural elements caused by live loads are computed by live load models defined in design codes. In most cases, such live load models are defined by studies performed on girder bridges, where extreme values of shear forces and bending moment are intended to be predicted. This paper shows that when code live load models are applied to truss bridges, the estimated forces in some structural elements may not be representative of those caused by actual traffic. Three WIM (weigh-in-motion) databases, which are recorded on roads in Mexico, are used as real traffic data. The results suggest that current code live load models are not entirely adequate to estimate forces in structural elements of truss bridges.

**Introduction**

Bridges are essential elements of the infrastructure in a region or country, which allow the communication and transportation of people and goods among different places, making them a necessary aspect in the daily life of any society. Due to conditions which bridges are subjected to during their useful life, the conceptualization, project, design, construction and maintenance of these structures are not free of uncertainties, which make bridge design a complex task that is still not fully understood. There are a wide variety of available materials and structural systems to build a bridge. One of the most used systems for medium and large spans are truss bridges, which in most cases are made of steel.

Structural elements in a truss bridge are mainly subjected to axial loads. Therefore, their behavior can be very different with respect to girder bridges, requiring specific studies to be performed to understand their overall behavior. Some studies have focused on damage identification, such as those carried out by Farhey *et al*. 1997 and Chang & Kim 2016, or structural health monitoring, like the one performed by Catbas *et al*. 2008.

Since structural elements in truss bridges have a specific way to transfer forces caused by vehicles, several studies have carried out experimental research aimed at improving the comprehension of truss bridges behavior, like those performed by Bakht and Jaeger 1990, Aktan *et al*. 1994, Azizinamini 2002, Häggström *et al*. 2014, and Täljsten *et al*. 2018. Likewise, Lee 1996, Wang *et al*. 2007, and Mulyadi & Suangga 2020 have conducted fatigue studies on truss bridges.

Other authors have focused on aspects that are more likely to occur in truss bridges, such as that carried out by Wang *et al*. 2017, who study the crack arising in a critical element of a three-span continuous truss bridge. Nichols *et. al* 2015 propose a repair scheme for the M4 Boston Manor viaduct due to weld defects.

Unfortunately, truss bridge collapses are also reported, such as the case presented by Stark *et al*. 2016, where an oversized container struck portions of the bridge causing the subsequent collapse. One of the most interesting studies in this context is carried out by Hao 2011 for the collapse of I-35W bridge, concluding that an inadequate sizing of gusset plates was the principal cause of collapse.

The work carried out by Parida and Talukdar 2020 is one of the few in which a clear distinction between the behavior of truss bridges and girder bridges is shown; they conclude that the dynamic amplification factor should not be a uniform value in each element of the bridge.

From the above-mentioned information, it can be inferred that truss bridges have their own characteristics, even more so if we consider the case of Mexico, where in late 1970s and early 1980s, a bridge typology called

Tridilosa became very popular. Tridilosa bridges are mainly composed of two parts: (1) a three-dimensional steel truss, and (2) a reinforced concrete slab over the truss. Figure 1 shows an example of Tridilosa bridge. To date, many Tridilosa bridges have exhibited severe fatigue problems and, in most cases, they have been replaced by prestressed concrete girder bridges. Nevertheless, a few Tridilosa bridges remain in operation, such as the one shown in Figure 1. Unlike Tridilosa bridges which tend to exhibit fatigue problems, girder bridges perform much better for practically the same conditions. This aspect is considered relevant by the authors. Consequently, one of the possible causes of this phenomenon is investigated in this study.

Code live load models have been proposed to represent load effects and forces generated by actual traffic during the bridge service life. It should be kept in mind that practically all code live load models have been proposed based on studies carried out on girder bridges, *i.e.,* they try to predict the extreme values of shear force and bending moments (and not necessarily full-length force envelopes) acting on girder bridges. Therefore, when such live load models are used to analyze truss bridges, the estimated axial force in some elements may be not entirely representative of that generated by the real traffic. In this study, this inconsistency by using a model of an existent truss bridge in Mexico, which configuration is more representative of common truss bridges (unlike the Tridilosa bridge mentioned before) is explored.

The main objective of the present study is to investigate possible inconsistencies between the forces in elements of truss bridges generated by code live load models and those generated by realistic traffic loads. Additionally, a possible reason of fatigue effects on a truss bridge configuration popular in Mexico in the 1970s and 1980s (and still existent) is investigated. To achieve these objectives, truss bridges considered in this paper are subjected to three WIM (weigh-in-motion) databases recorded on highways in central Mexico, and the computed acting forces are compared with those generated by live load models from different bridge codes, such as the Mexican IMT/SCT 1999 and 2004, the AASHTO 2017, as well the Canadian CAN/CSA 2014.

**Types of bridge analyzed**

Two types of truss bridges are considered to show the inconsistencies of using code live load models in this paper. The first one is an existent truss bridge named "*El Infiernillo*" (The little hell), located in Mexican federal 37 (toll) highway, between Guerrero and Michoacán states in the southwest of Mexico. The total length is 360 m divided in 8 spans, three at each extreme of the bridge having an approximate length of 26 m, and other two central spans with a length of 102 m. The extreme spans are built of prestressed concrete girders; the central spans correspond to camelback trusses, as schematically shown in Figure 2. The bridge allows two-lane traffic (one lane for each direction) crossing the reservoir with a dam of the same name ("*El Infiernillo*"), a main hydroelectric facility in Mexico given its power generation capacity. The main trusses are connected with by transversal floor trusses in the lower nodes. The traversal trusses support a concrete slab over steel stringers, forming the composite system over which the asphalt road surface in placed. As shown in Figure 2, the geometry of the main trusses of "*El Infiernillo*" bridge is similar to that of many other bridges. Consequently, the conclusions could be applicable to other truss bridges.

Additionally, two Tridilosa-type bridges with 9 m and 18 m long spans, like the ones schematically showed in Figure 3, are used to inspect the inconsistencies of employing code live load models to obtain element forces in these bridges. This is also aimed at exploring whether induced forces on structural elements caused by code live load models, in comparison to those by actual traffic, are one of the main reasons why these bridges are more susceptible to fatigue issues.

**Code live load models**

Several live load models defined in different bridge design codes are used to predict the element forces in the studied truss bridges. T3-S3 (Figure 4a) and T3-S2-R4 (Figure 4b) are considered (IMT/SCT 1999) since these live load models were widely used to design Tridilosa-type bridges; they correspond to actual truck configurations in Mexico. Also, the T3-S2-R4 model was used to design *"El Infiernillo"* bridge. The IMT 66.5 (Figure 4c) defined in IMT/SCT 2004 is also considered, which was developed based on traffic surveys carried out in the 1990s in Mexico. Also, the HL-93 (Figure 4d) and CL-925 (Figure 4e) models defined by the American (AASHTO 2017) and Canadian (CAN/CSA 2014) standards, respectively, are used.

It is noted that the goal of using the different models above is not to rank the different models mentioned above, *i.e.*, it is not intended to define which model is better, since they were developed for different vehicular traffic characteristics. The aim is focused in evaluating the ability of the models to predict the forces

(tensions and compressions forces, as well their variation ranges) in truss bridge elements due to traffic loading, since all models are derived from studies essentially based on girder bridges, where the maximum values of shear force and bending moment are the main interest (not axial loads).

**WIM databases**

Three WIM databases for bridges analyses are used. The first one has been used in previous works (García–Soto *et al*. 2015 and 2020) and was recorded in a four-lane highway connecting the cities of Irapuato and La Piedad (45 federal highway in central Mexico) during January to March 2009 encompassing 3,834,961 vehicles recorded by permanent WIM stations. Figure 5a shows the histogram for all vehicles in this database according to their Gross Vehicular Weight (GVW), while Figure 5b shows histogram for vehicles with GVW > $P_{90}$(GVW), *i.e.*, the distribution of 10% heaviest vehicles in the database. For convenience these WIM data will be identified hereinafter as IP–2009.

The second database was recorded in the same highway for one week (January $24^{th}$ – $30^{th}$, 2017) encompassing 226,488 vehicles recorded by temporal WIM stations. Figure 5c shows the histograms distribution for all vehicles according to their GVW. As in the previous case, Figure 5d shows the vehicle distribution with GVW > $P_{90}$(GVW). These WIM data will be defined hereinafter as IP–2017.

The third database was recorded in a two-lane road connecting the cities of Guanajuato and Silao (central Mexico, 110 federal road) for one week (January $16^{th}$ – $22^{nd}$, 2017) encompassing 127,218 vehicular data recorded also by a temporal WIM equipment. Figures 5e and 5f shows the GVW data distribution in histograms in the same way as previously presented. As can be noted, being a highway which allows lighter traffic, the vehicle weights are considerably lower than those for the above-mentioned databases. As in the previous cases, for convenience these WIM data will be identified hereinafter as GS–2017.

It is noted that a screening was carried out in the three databases to discard erroneous data, discarding vehicles with zero forces at any axle, and a few cases with GVW greater than 130 metric ton (1,274.86 kN), as they are considered instrumental errors in data acquisition.

**Software and Analyses**

To perform all required analyses the AMER 2.0 software developed by Hernández–Martínez 2019 is used. This software has been used for determining load effects in continuous span bridges (García-Soto et al. 2020) and allows computing extreme forces on elements induced by each vehicle from the databases referred to earlier, as well as by the code live load models considered (Figure 4). All analyses are performed by running individually every vehicle over the truss bridges with 10 cm increments in both directions, because vehicular loads are not symmetrical about their mass center, although bridges geometries are symmetrical. It is highlighted that one of the main characteristics of the AMER 2.0 is its ability to efficiently analyze moving loads over bridges, which allows to carry out the very time-consuming task of running millions of vehicles, in small increments and both directions, in a reasonable time, which is very unlikely to be achieved by using commercial software. Both bridge types are modeled as trusses with pinned connections using beams elements to transfer vehicle loads to truss nodes, which has been used by O'Connell and Dexter 2001 who found a good correlation with respect to experimental results. Loads are transferred to nodes at the top chord for Tridilosa-type bridges and to the bottom chord for *"El Infiernillo"* bridge.

**Analysis of Results**

A good live load model should adequately estimate force magnitudes in all bridge elements during its service life. As the magnitudes and frequencies of actual live loads evolve on time, it is not surprising that estimated force magnitudes from code live load models are exceeded by actual vehicles; however, the number of times which estimated forces from a code live load model are exceeded by traffic should remain approximately constant in all bridge elements, which is shown in Figure 6 for the top chord of "*El Infiernillo*" bridge.

The left side of Figure 6 shows the forces acting on top chord elements along the bridge for IP-2009 (Figure 6a), IP-2017 (Figure 6c) and GS-2017 (Figure 6e). The convention that compression and tension forces are negative and positive, respectively, is adopted. A scale of gray colors is used to represent the force distribution acting on each member along the bridge for the used databases; lighter gray (15% gray) is used to represent the range in which the minimum and maximum values of acting forces on elements along the bridge are presented. Lower and upper limits of gray at 25% are used to represent the 1% and 99% of the largest forces, (*i.e.*, $P_1$ and $P_{99}$ respectively; this convention is adopted for other percentages). In the same way, lower and upper limits of gray at 40% represent the $P_5$ and $P_{95}$ force value distribution. $P_{10}$ and $P_{90}$ force value distribution are represented by lower and upper limits of darkest gray (65% gray). Forces distributions are on the negative side because top chord elements are subjected to compression forces only. Since limits for lower value forces can be very close, the way of presenting the results for each database allows a better visualization of the largest force values acting on elements. Figures 6a, 6c and 6e make clear the fact that only a very small fraction of the heaviest trucks causes the largest forces in the chord; although this could be expected, the figures allow to inspect this issue at a glimpse. The force values caused by each code live load model considered are plotted with markers. All presented results are unfactored. The study is neither intended to discuss the adequacy of each model to capture the actual forces from the databases, nor to be calibrated to represent truss forces for given return periods (extrapolation or analysis of extremes is out of the scope of this study). The use of code live load models is rather envisaged to inspect whether the models approximately capture the shape of the real element forces; if this is possible, a calibration task to reach desired extrapolated values will be feasible. In fact, it can be observed in Figures 6a, 6c and 6e that all the models adequately follow the trend of the maximum compressions (for the top chord). We will come back to this aspect for other truss elements later.

The right side of Figure 6 illustrates the number of times (in percentage) that vehicle induced forces from the database exceed the value for each live load model, *i.e.*, the exceedance rates for structural elements along the bridge are shown; Figure 6b for IP-2009, Figure 6d for IP-2017 and Figure 6f for GS-2017. As can be noted, exceedance rates are practically uniform for top chord elements for all code live load models, which can be associated with a uniform probability of failure for those bars. Thus, if exceedance rates would be too large to reach a selected reliability level, a recalibration of live load factors for design force combinations is expected to lead to uniform reliabilities, for all truss elements along the top chord of the bridge (for all code models considered). As can be seen in Figures 6b, 6d and 6f, the exceedance rate values are directly related to the GVW in each database, leading to null exceedance rates for GS-2017, since in this database there are only light vehicles, which do not generate greater forces than code live load models. It becomes evident that different live load models (and/or live load factors) should be prescribed for highways with different traffic characteristics. Similar results for the bottom chord of *"El Infiernillo"* bridge are observed for all databases considered, except that tensions instead of compressions are obtained, and they are not shown for brevity.

To draw a parallel between *"El infiernillo"* truss bridge and a simply supported girder bridge of the same span (*i.e.,* 102 m) for exceedance rates, it is noted that results for shear force and, specially, for bending moment are practically the same as those shown in Figure 6. This is because code live load models are calibrated from load effects for girder bridges, aimed at correctly estimating the values of shear force and bending moment. It should be also noted that acting forces at bottom and top chords are, in fact, an equivalent form of bending moment.

So far, no relevant inconsistencies in the ability of truck models to capture (at least in relation to the trends shape) the realistic truss element forces are observed. However, Figure 7 illustrates results for vertical truss elements in *"El infiernillo"* bridge and, unlike top chord elements, which are subjected to compression forces only, vertical elements are subjected to load reversals, so two force distributions are shown on the left-hand side of Figure 7 (for the three WIM datasets, analogously to Figure 6), positive for tension forces, negative for compression. For these vertical truss elements, the exceedance rates are no longer uniform along the bridge; moreover, for IMT-66.5 and HL-93 models, there are elements with 100% of exceedance rates (as shown in the right-hand side of Figure 7). This occurs because those live load models include uniformly distributed loads, leading to some members to be always in tension or compression, being these models unable to adequately represent the load reversals that occurs when running the vehicles in the databases along the bridge. Similar trends as those in Figure 7 are found for the diagonal truss elements and are not shown for brevity. Overall, Figure 7 shows that code live load models do not capture the range variation of forces in vertical elements (or diagonals) as uniformly as in chord elements (differences are not discussed in detail for each model but they can be inspected in the figure). Therefore, a not so uniform reliability level (Garcia-Soto et al., 2015 and 2020) would be expected for different bridge truss elements. It is considered that there is room for improvement of current available code live load models by using optimization techniques with information like the one presented in Figures 6 and 7, and even to propose live load models specifically developed for truss bridges, given the obtained results and the fact that something similar was found for continuous bridges (Garcia-Soto et al., 2020) and the results for the Tridilosa-type bridges discussed in the following.

Figures 8 and 9 show the results for bottom diagonals for Tridilosa-type bridges of 9 m and 18 m span length, respectively. As can be noted, the following elements to those connected at supports (*i.e.,* second and penultimate diagonal members) have the highest exceedance rates for all databases. It should also be noted that highest exceedance rates are not related directly with GVW as in the top chord of *"El Infiernillo"* bridge. This may occur since vehicles in IP-2017 and GS-2017 databases have higher loads per axle, although the GVW is not necessarily heavier. In addition, these cases correspond to short bridges, where heavier and longer vehicles do not entirely fit on these bridges.

**Discussion**

The fact that structural elements exhibit non-uniform exceedance rates along the bridge (Figures 7 to 9) implies that elements with higher exceedance rates have a greater probability of excessive deformations, fatigue, cracks, etc., *i.e.*, they are expected to have more critical conditions associated with different limit states.

It is acknowledged that results showed here correspond only to live loads. A more comprehensive study, recommended for future research, should include forces from different design load combinations, traffic scenarios, dynamic loading, etc. Nonetheless, results may indicate that some problems, which have occurred in existent bridges, could be caused (at least in part) due to the different (implicit) safety levels that different truss elements are designed to using existent live load models. Empirical evidence could be indicative of this problem. For instance, *"El Infiernillo"* bridge has been retrofitted several times; Tridilosa-type bridges have even been replaced or, as in is the case of the bridge in Figure 1, additional elements had to be added to diagonal elements near the supports, in order to reduce their slenderness ratio to increase their compressive strength.

If instead of the analyzed bridges, girder bridges with the same span lengths of 102, 18 and 9 m , are considered, even though not perfect uniform exceedance rates for shear force and bending moment are found, they are much more uniform than those presented in Figures 7 to 9 (*i.e.*, for girder bridges results resembles those of Figure 6). These inconsistencies seem to indicate that using live load models derived from girder bridges may not be entirely adequate for estimating acting forces on truss bridges, especially for vertical and diagonal truss elements, rather than for the chords. Failures and issues reported in the literature may be due, at least in part, to this inability of the existent code live load models to predict uniform demands for all the truss elements. For instance, the I-35 bridge collapse is attributed to a lack of consideration of the forces from diagonal truss members and to the fact that traffic loads, at the collapse time, led to the highest stresses in the failed gusset plates (those connected to diagonal elements), which should have been twice as thick (Hao 2010). Other studies have also reported problems in diagonal or vertical members associated to failures or underperformance due to wind loading and fatigue loading but also to experimental live loading (*e.g.*, Wang *et al.*, 2007; Wang *et al.*, 2017; Azizinamini, 2002). Although these issues are not always directly associated with live load effects, the findings in the present study may indicate that this could be a contributing (unnoticed) factor leading to failures and should be thoroughly investigated in future projects.

Moreover, if this is confirmed by extensive studies on different truss bridge configurations of varying span lengths, bridges with other structural types, such as frame bridges, suspension bridges, cable-stayed bridges, arch bridges, etc., could also exhibit a similar problem, *i.e.,* current code live load models, developed mainly for girder bridges and simply supported spans, may not be able to represent uniform demand levels in all bridge components. Such future research is recommended, since it may help to answer problems and failures that are not yet fully understood. It also could lead to the development of code live load models for designing bridge elements for uniform structural reliability levels.

Although differences are expected for other WIM data, because different countries may have different traffic populations, the results in the present study may hold in general terms. Similar studies should be conducted using other WIM databases to inspect whether similar trends are found, leading to similar conclusions.

## Conclusions

In this study millions of vehicles from three WIM (weigh-in-motion) databases are run over realistic truss bridge models based on existent truss bridges in Mexico. Each vehicle is run along the bridge in both directions in 10 cm increments. Such a computing challenge is achieved by using a software expressively developed by one of the authors to efficiently compute forces in truss elements. To the authors knowledge, this is the first time that such an extensive study in carried out. Resulting forces are compared versus those generated by code live load models from different countries.

Currently, code live load models, used to estimate acting forces caused by actual traffic on bridge elements, are mostly based on studies for girder bridges. Results presented in this paper suggest that such models are not entirely suitable for estimating forces that may be present in truss bridge elements, being the differences in some cases significant, in terms of uniform demands for all truss elements.

The exceedance rates (*i.e*., percentage of element forces) in truss bridges generated by actual traffic, with respect to forces estimated by code live load models, can be very irregular along the bridge for some structural elements. This is especially true for elements that exhibit load reversals, such as vertical and diagonal elements with considerable force magnitudes. Non-uniform exceedance rates imply a poor estimation of similar demands (acting forces) for every bridge structural element. Failure to properly estimate acting forces in bridge elements may lead to designing truss elements with different (implicit) safety levels (*e.g.,* in terms of reliability), which in turn could make some of them more susceptible to resistance, deformation and fatigue problems. An important aspect for the fatigue problem is not only the range of stress variation but also the magnitude of stresses at which this variation occurs. Thus, a deficient estimation of acting forces on bridge elements could make the emergence of fatigue problems more likely. The failures in diagonal elements in truss bridges reported in the literature could support this conclusion, since the found inconsistencies between real element forces and those generated by existent code live load models may be a contributing (unnoticed) factor leading to such failures.

For truss bridges with some lengths, results indicate that models with uniformly distributed loads may not be adequate to represent load reversals, especially for elements near supports, as they tend to only exhibit compressive or tensile forces, due the presence of the uniform loading.

Consequently, evaluating all critical cases that can occur for live load effects with only one live load model for all bridge types may be not the most suitable approach. Thus, it seems feasible that design codes should have a set of live load models, where each model focuses on eliciting one/some critical conditions for bridge design/evaluation.

With the aspects presented in this article, the authors intend to respond to Hao's call (Hao 2010) for "more prominent discussions and shed light on the potential safety issues that may exist in the truss bridges that are currently in service."

**Data Availability Statement**

Databases used to support the findings of this study are confidential in nature and may only be provided with restrictions. IP-2009 database is previously by García-Soto *et. al.* 2015 and 2020. IP-2017 and GS-2017 databases are expected to be published soon as data article. AMER 2.0 software are free to download at http://www.di.ugto.mx/GEMEC/ site. An updated software version of AMER 2.0 is available from the corresponding author upon reasonable request.

**Acknowledgements**

The financial support from CONACyT (National Research and Technology Council of Mexico; Project "Problemas Nacionales 2014" No. 248162) to get IP-2017 and GS-2017 WIM databases is gratefully acknowledged.

**References**

AASHTO (American Association of State Highway and Transportation Officials) 2017. Bridge Design Specifications. Washington, DC. AASHTO.

Aktan, A. E.; Lee, K. L.; Naghavi R.; and Hebbar K. 1994. "Destructive testing of two 80-year-old truss bridges." Transp. Res. Rec. 1460, 62-72. http://onlinepubs.trb.org/Onlinepubs/trr/1994/1460/1460-008.pdf.

Azizinamini A. 2002. "Full scale testing of old steel truss bridge." J. Constr. Steel Res. 58(5-8), 843-858. https://doi.org/10.1016/S0143-974X(01)00096-7.

Bakht, B. and Jaeger, L. 1990. "Bridge testing – A surprise every time." J. Struct. Eng. 116(5), 1370-1383. https://doi.org/10.1061/(ASCE)0733-9445(1990)116:5(1370).

CAN/CSA (Canadian Highway Bridge Design Code) 2014. Canadian Highway Bridge Design Code. Canadian Standards Association. Mississauga, Ontario, Canada. CAN/CSA.

Catbas F. N.; Susoy M.; and Frangopol D. M. 2008. "Structural health monitoring and reliability estimation: Long span truss bridge application with environmental monitoring data." Eng. Struc. 30(9), 2347-2359. https://doi.org/10.1016/j.engstruct.2008.01.013.

Chang, K. C. and Kim C. W. 2016. "Modal-parameter identification and vibration-based damage detection of a damage steel truss bridge." Eng. Struct. 122(1), 156-173. https://doi.org/10.1016/j.engstruct.2016.04.057.

Farhey, D. N.; Thakur A. M.; Buchanan R C; Aktan A. E; and Jayaraman N. 1997. "Structural deterioration assessment for steel bridges." J. Bridge Eng. 2(3), 116-124. https://doi.org/10.1061/(ASCE)1084-0702(1997)2:3(116).

Garcia–Soto, A. D.; Hernández–Martínez A.; and Valdés–Vázquez J. G. 2015. "Probabilistic assessment of a design truck model and live load factor from weigh-in-motion data for Mexican highway bridge design". Can. J. Civ. Eng. 24(11). https://doi.org/10.1139/cjce-2015-0216.

García–Soto, A. D.; Hernández–Martínez A.; and Valdés–Vázquez J. G. 2020 "Probabilistic assessment of live load effects on continuous span bridges with regular and irregular configurations and its design implications". Can. J. Civ. Eng. 24(4). https://doi.org/10.1139/cjce-2018-0232.

Hao, S. 2010. "I-35W bridge collapse." J. Bridge Eng. 15(5). https://doi.org/10.1061/(ASCE)BE.1943-5592.0000090.

Häggström, J; Collin, P.; Blanksvärd, T.; and Täljsten B. 2014. "Assessment and full sacale failure test of a steel truss bridge". 37[th] IABSE Madrid Symposium Report. 102, 2757-4764. http://ltu.diva-portal.org/

Hernández–Martínez A. 2019. "Análisis Matricial de Estructuras Reticulares" AMER 2.0 [Matrix Analysis of Reticular Structures]. Available free at http://www.di.ugto.mx/GEMEC/

IMT/SCT (Instituto Mexicano del Transporte/Secretaría de Comunicaciones y Transportes [Mexican Institute of Transportation/Ministry of Communications and Transportation]). 1999. "Modelo de cargas vivas vehiculares para el diseño estructural de puentes en México. [Vehicular live load model for structural design of bridges in Mexico]" Instituto Mexicano del Transporte y Secretaría de Comunicaciones y Transportes. Sanfandila, Querétaro, México. IMT/SCT [In Spanish]. https://www.imt.mx/archivos/Publicaciones/PublicacionTecnica/pt118.pdf

IMT/SCT (Instituto Mexicano del Transporte/Secretaría de Comunicaciones y Transportes [Mexican Institute of Transportation/Ministry of Communications and Transportation]). 2004. "Formulación de la norma SCT de cargas vehiculares para el diseño estructural de puentes carreteros. [Formulation of SCT standard of vehicular loads for structural design of highway bridges]" Instituto Mexicano del Transporte y Secretaría de Comunicaciones y Transportes. Sanfandila, Querétaro, Mexico. IMT/SCT [In Spanish]. https://imt.mx/archivos/Publicaciones/PublicacionTecnica/pt243.pdf

Lee, S. B. 1996. "Fatigue failure of welded vertical members of a steel truss bridge." Eng. Fail. Anal. 3(2), 103-108. https://doi.org/10.1016/1350-6307(96)00003-9.

Mulyadi, E. and Suangga M. 2020. "WIM data analysis for the fatigue lifetime evaluation of standard steel truss bridge elements." IOP Conf. Ser.: Mater. Sci. Eng. 1007 012155. doi:10.1088/1757-899X/1007/1/012155.

O'Connel, H. M. and Dexter, R. J. 2001. "Response and analysis of steel trusses for fatigue truck loading." J. Bridge Eng. 6(6), 628-638. https://doi.org/10.1061/(ASCE)1084-0702(2001)6:6(628).

Parida, S. and Talukdar S. 2020. "An insight to the dynamic amplification factor for steel truss girder bridge." Int. J. Steel Struct. 20, 1341-1354. https://doi.org/10.1007/s13296-020-00364-y.

Stark, T. D.; Benekohal, R.; Fahnestock, L. A.; LaFave, J. M.; He, J.; and Wittenkeller C. 2016. "I-5 Sakagit River Bridge Collapse Review." J. Perform. Constr. Facil. 30(6). https://doi.org/10.1061/(ASCE)CF.1943-5509.0000913.

Täljsten, B.; Blanksvärd, T.; Sas, G.; Bagge, N.; Nilimaa, J.; Popescu, C.; and Elfgren, L. 2018. "Bridges tested to failure in Sweden." IASBE Conference Copenhagen 2018. http://ltu.diva-portal.org/.

Wang, C. S.; Qian, H.; Zhan, A.; Xu, Y.; and Hu, D. 2007 "Fatigue and fracture evaluation of a 70 year old steel bridge." Key Eng. Mater. 347, 359-364. https://doi.org/10.4028/www.scientific.net/KEM.347.359.

Wang, Q.; Nakamura, S.; Okumatsu, T,; and Nishikawa, T. 2017. "Comprehensive investigation on the cause of a critical crack found in a diagonal member of a steel truss bridge." Eng. Struct. 132(1), 659-670. https://doi.org/10.1016/j.engstruct.2016.11.049.

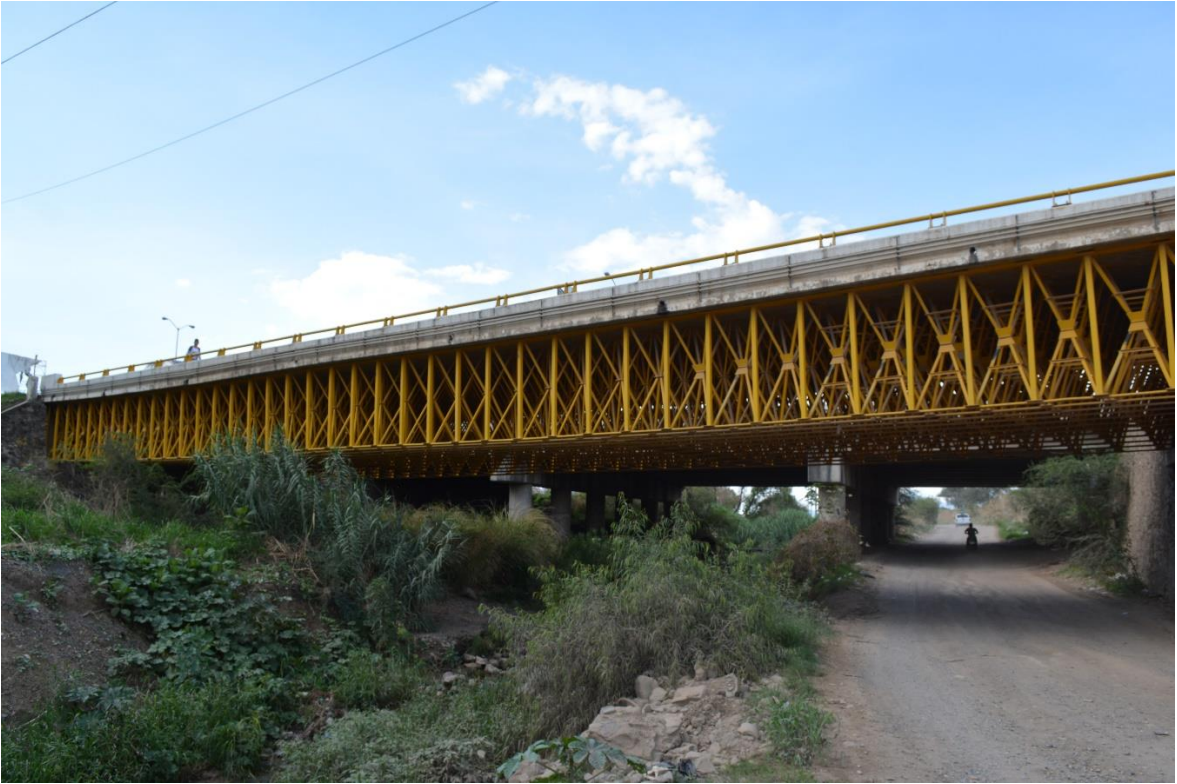

Figure 1 Tridilosa Bridge

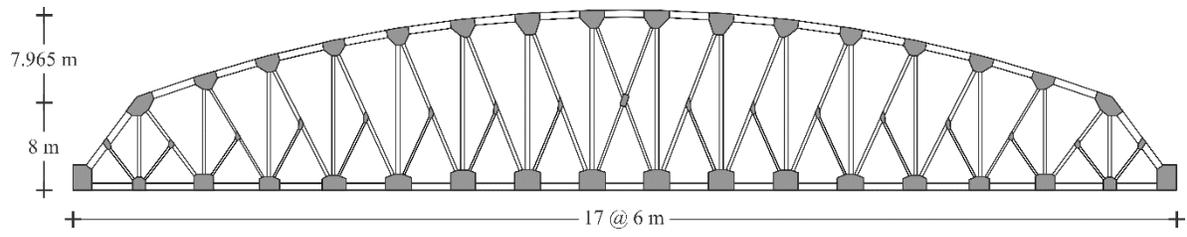

Figure 2 Main truss of *"El Infiernillo"* bridge

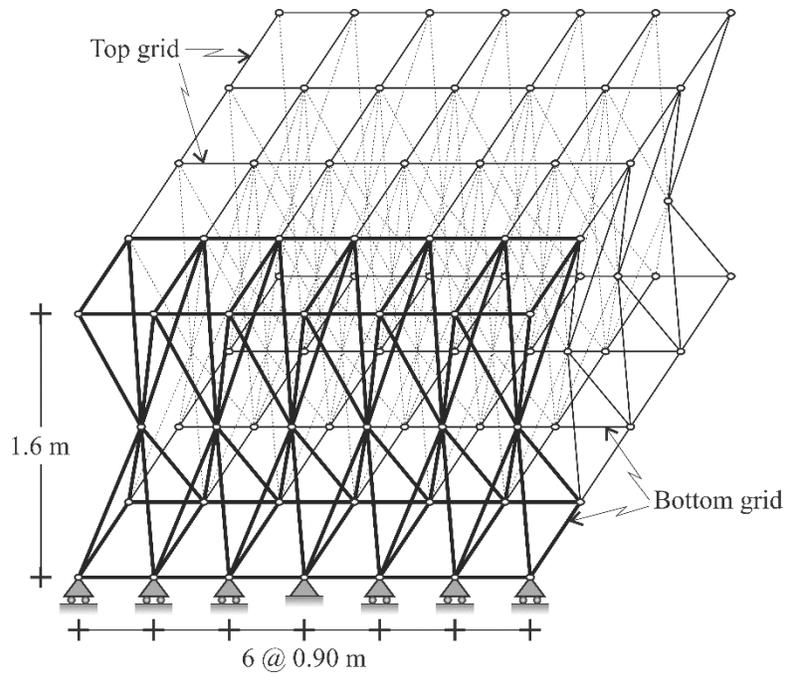

a) Isometric transversal view

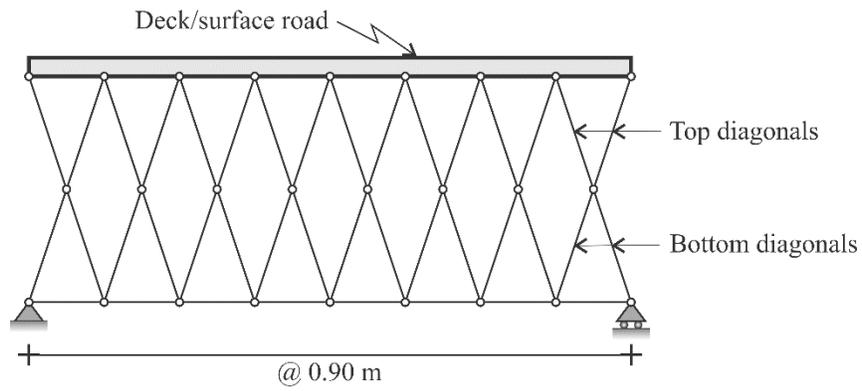

b) Longitudinal view

Figure 3 Geometry of Tridilosa-type bridges

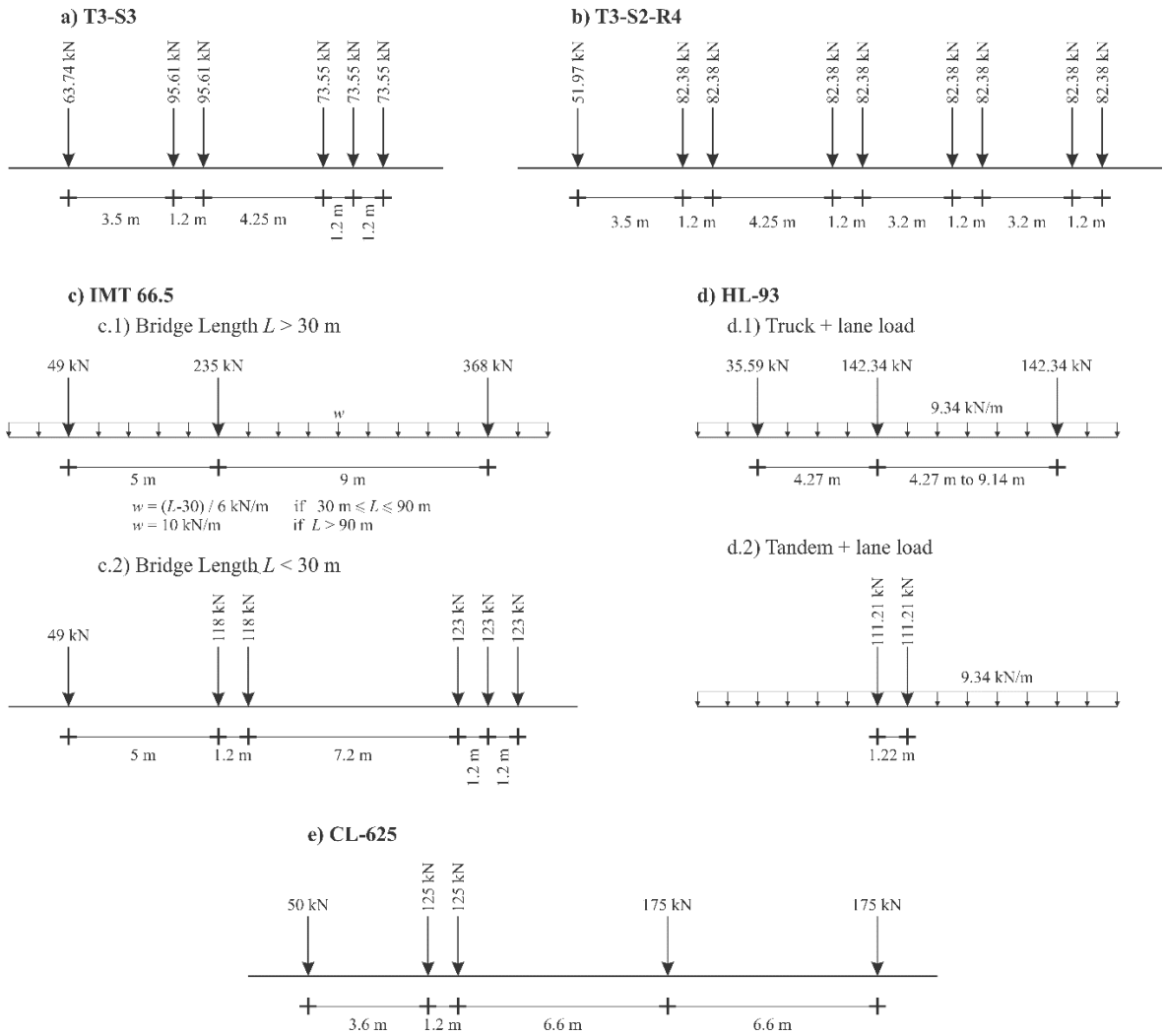

Figure 4 Code live load models

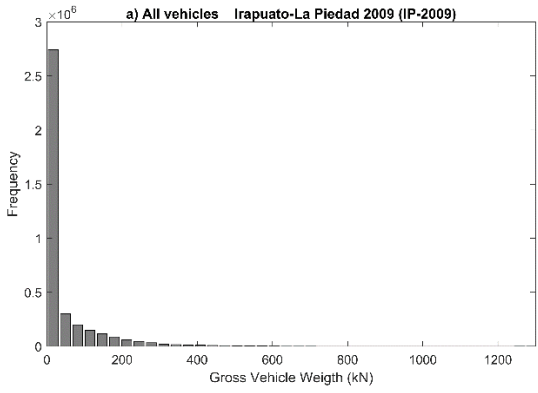
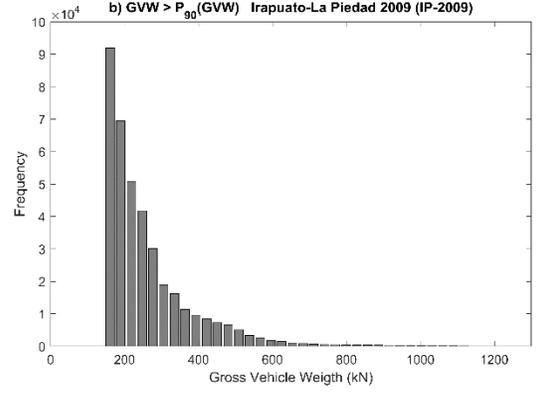
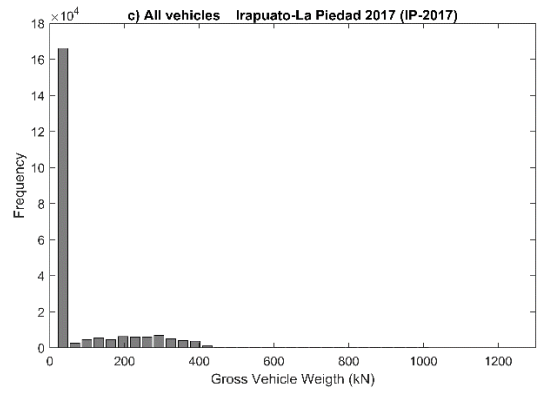
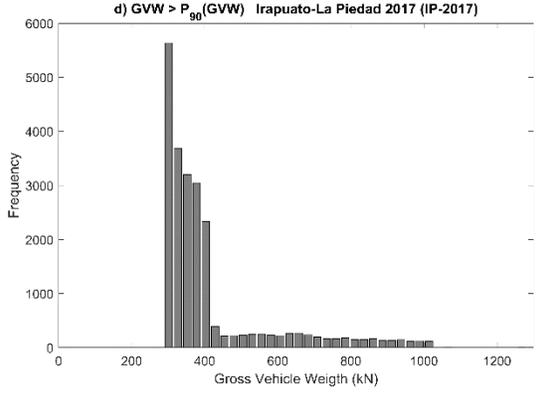
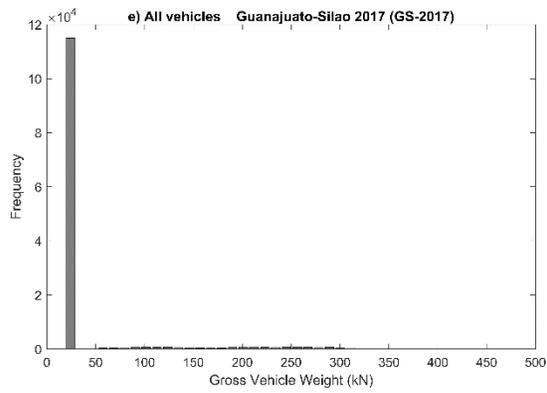
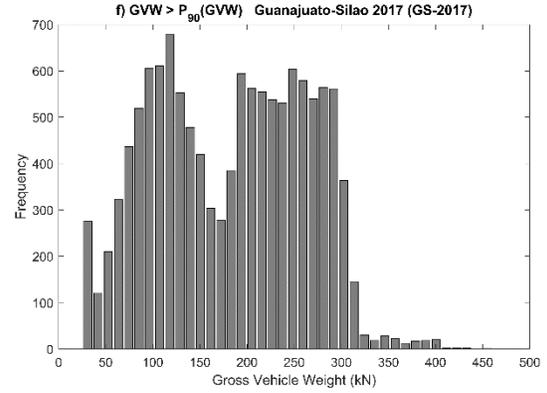

Figure 5 GVW distribution for used WIM databases

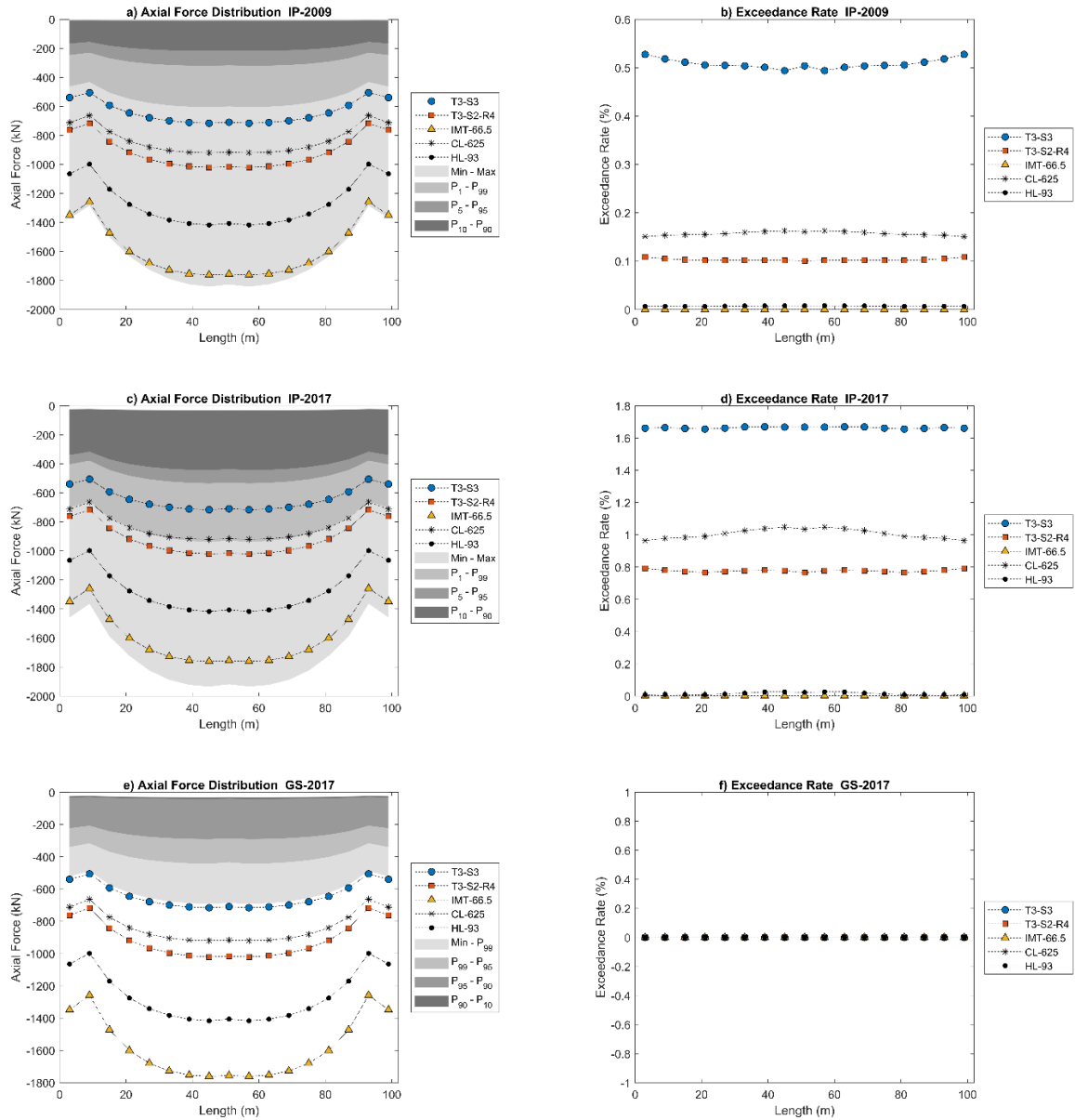

Figure 6 Results for top chord of *"El Infiernillo"* bridge

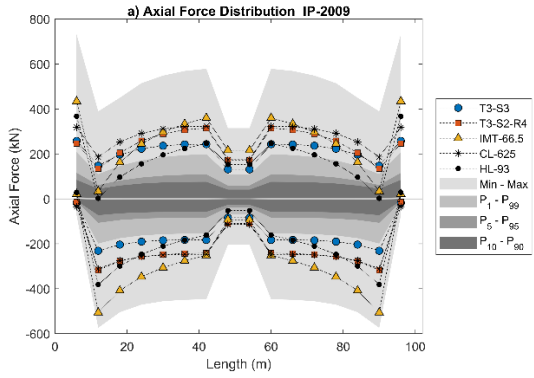
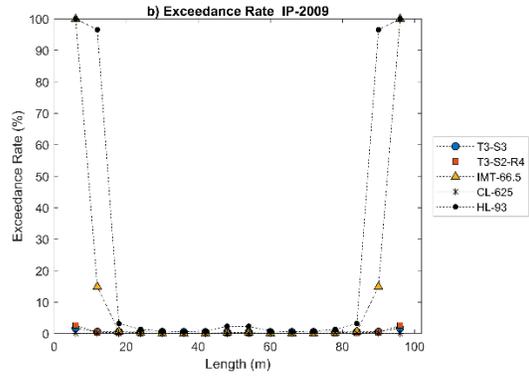
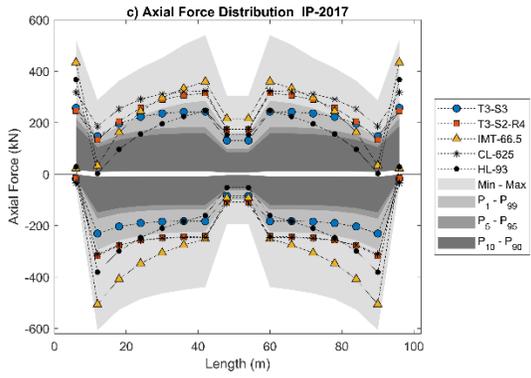
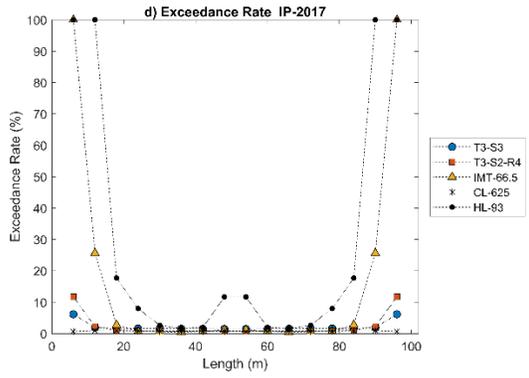
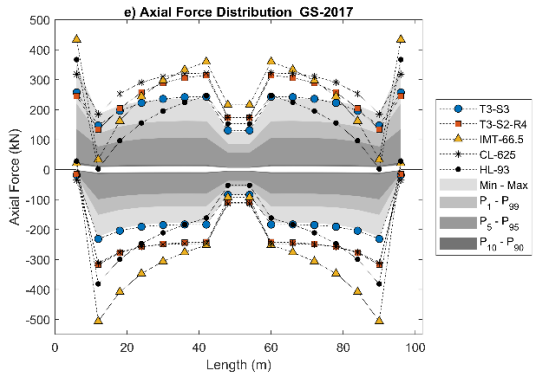
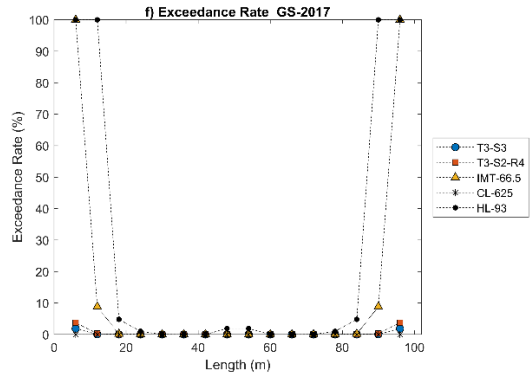

Figure 7 Results for vertical elements of *"El Infiernillo"* bridge

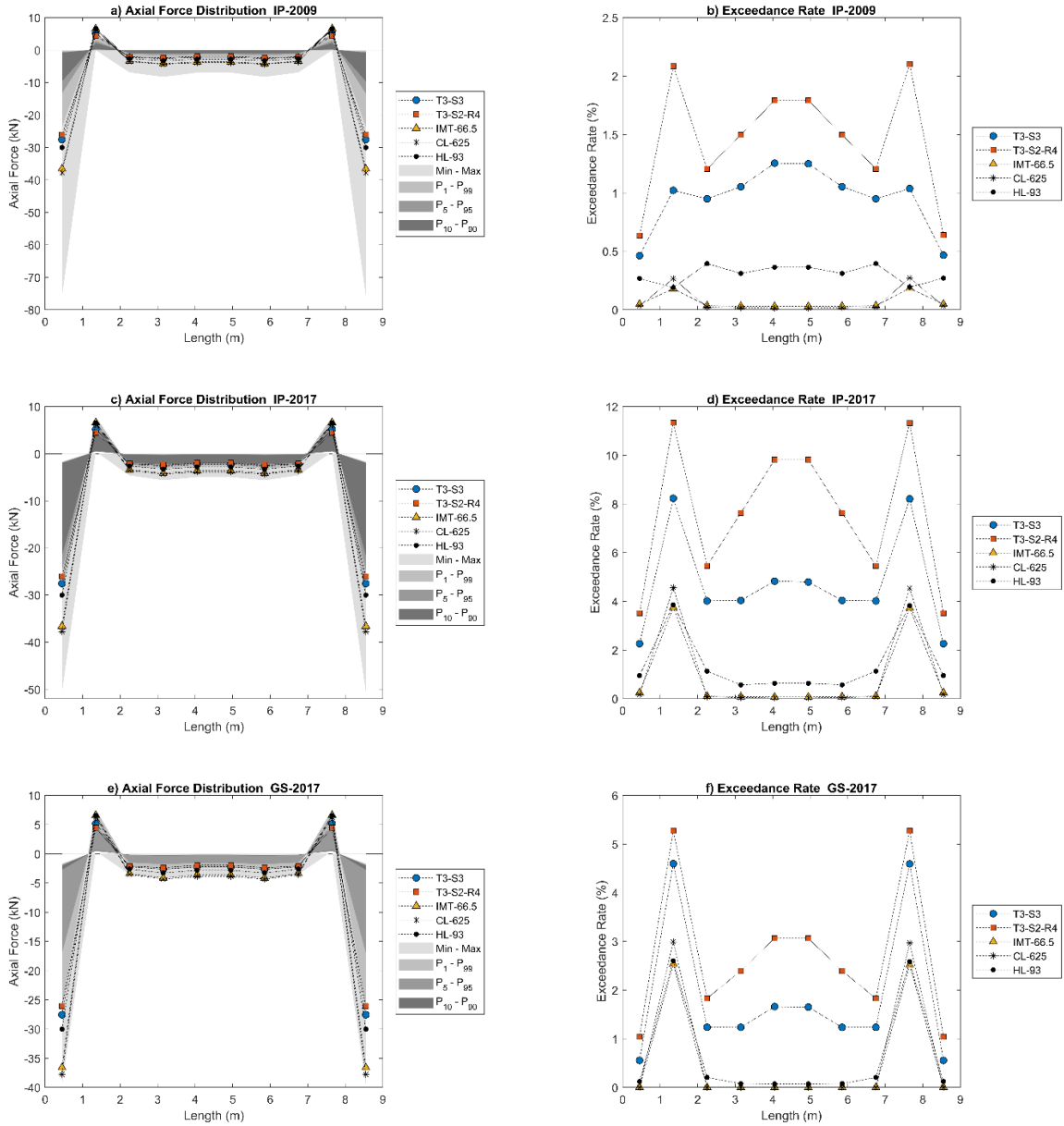

Figure 8 Results for bottom diagonals for 9 m long Tridilosa-type bridge

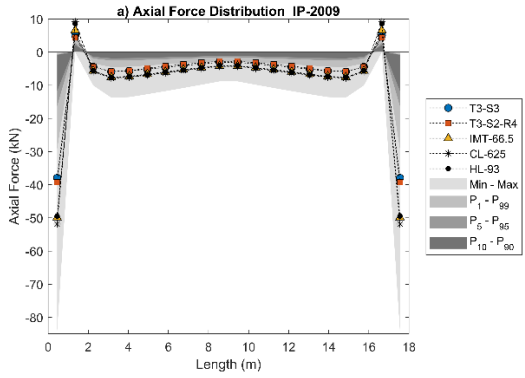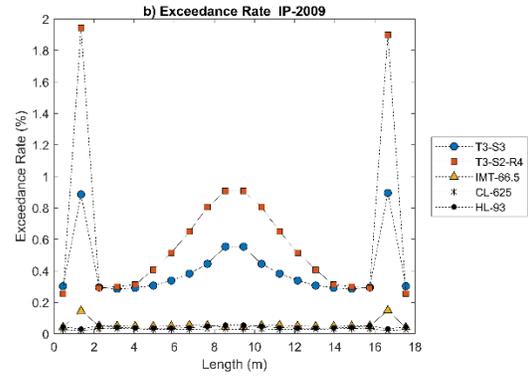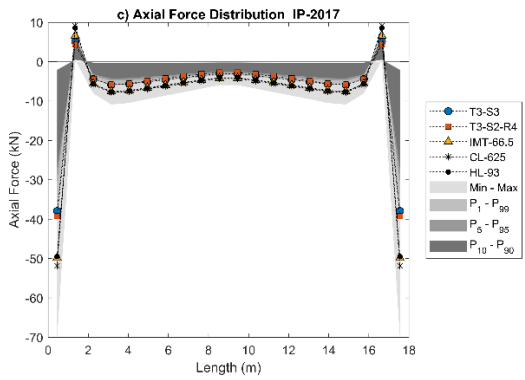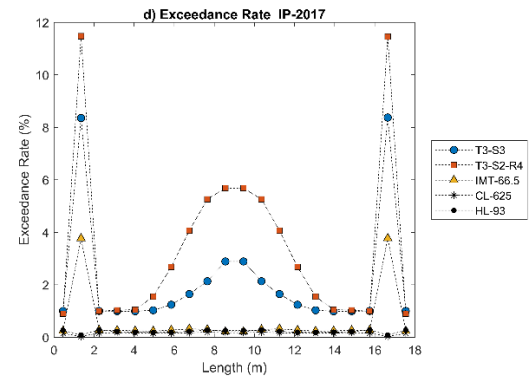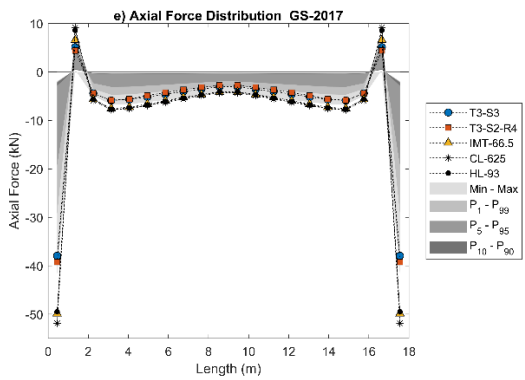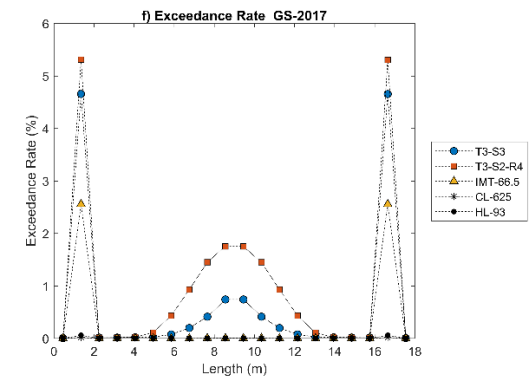

Figure 9 Results for bottom diagonals for 18 m long Tridilosa-type bridge